\documentclass[11pt,twoside]{article}

\usepackage{prcsa2008}
\usepackage{epsf}
\usepackage{epsfig}
\usepackage{lscape}

\begin{document}
\title{Non-Linear Cepheid Period-Luminosity Relation and the Interaction of Stellar Photosphere with Hydrogen Ionization Front}   
\author{C. Ngeow}   
\affil{University of Illinois, Urbana, IL 61801 USA}    

\author{S. M. Kanbur}   
\affil{State University of New York at Oswego, Oswego, NY 13126 USA}    

\begin{abstract} 

The Cepheid period-luminosity (P-L) relation is regarded as a linear relation (in log[P]) for a wide period range from $\sim2$ to $\sim100$ days. However, several recent controversial works have suggested that the P-L relation derived from the Large Magellanic Cloud (LMC) Cepheids exhibits a non-linear feature with a break period around 10 days. Here we review the evidence for linear/non-linear P-L relations from optical to near infrared bands. We offer a possible theoretical explanation to account for the nonlinear P-L relation from the idea of stellar photosphere - hydrogen ionization front interaction. 

\end{abstract}

\section{Introduction}

The Cepheid period-luminosity (P-L) relation takes the form of $M_{\lambda}=a_{\lambda}\log (P) + b_{\lambda}$, where $a$ and $b$ are the slope and zero-point (ZP) for the P-L relation at a given $\lambda$ bandpass, respectively, and $P$ is the pulsation period in days. The Cepheid P-L relation is the first rung of the distance ladder and has been widely applied to the distance scale studies, which deliver less than 10\% accuracy in distance \citep[for example, see][]{fre01}. The Cepheid P-L relation can also be used in the stellar pulsation and evolution studies, by comparing the theoretical P-L relations to empirical results. The Cepheid P-L relation has been assumed to be \emph{linear} and \emph{universal} for a long time. Hence Cepheids in the Large Magellanic Cloud (LMC) were used to derive the P-L relation due to its nearby proximity and the presence of a large number equal-distant Cepheids \citep[for example, see][]{uda99}. 

Nearly a century after its discovery by Leavitt in 1912, there are still some open questions regarding the Cepheid P-L relation. One of them being the metallicity dependency of the ZP. The latest empirical determination of this dependency is: $\gamma=-0.29\pm0.09(R)\pm0.05(S)$mag dex$^{-1}$ \citep{mac06}. Furthermore, recent works have strongly suggested that the Cepheid P-L relation could be neither universal \citep*{tam03,nge04,fio07} nor linear \citep*{kan04,kan06,san04,nge05,nge06a}, in contradiction to the usual assumptions for the Cepheid P-L relation. In this Proceeding, we concentrate on the discussion of the non-linear P-L relations as observed from the LMC Cepheids. 

\section{The Non-Linear LMC Cepheid P-L Relation}

Even though almost all of the distance scale studies assume that the LMC P-L relation is linear, high quality data for a large number ($\sim10^3$) of fundamental-mode LMC Cepheids have provided evidence to imply that the optical LMC P-L relations are \emph{not} linear, with a discontinuity in the slope seen around $P=10$ days (see references given in the Introduction). The justification and the significance of selecting the break period at 10 days has been discussed and given in the same references above. Various tests have been performed to look for the causes of this non-linearity, which include observing strategies, photometric errors, extinction errors, influence of outliers, number of long period Cepheids in the samples and contamination by overtone Cepheids. However, none of these or any combination of them is found to be responsible for the observed non-linear LMC P-L relation \citep{kan04,kan06,san04,nge05,nge06b}. This is in contradiction to the existing paradigm that the Cepheid P-L relation should be linear. 

Due to intrinsic dispersion along the P-L relation (roughly $\sim0.2$mag in $V$ band) caused by the finite width of the instability strip, the non-linearity of the P-L relation is difficult to visualize. This is convincingly demonstrated from Figure 1 of \citet{nge06b}, which presents two simulated P-L relations constructed from an intrinsicly linear and non-linear P-L relations, respectively. Therefore, careful \emph{statistical analyses} are required to detect the presence of non-linearity in the P-L relation \citep{nge06b}. These have included the $F$-test, non-parametric methods such as LOESS, regression methods robust to outliers, the testimator approach and Bayesian Information Criterion methods. All have supported the existence of the non-linear LMC P-L relation at more than a 95\% confidence level and a ``break'' period around 10 days \citep{kan04,kan06,nge05,kan07,nge08a}.

\begin{table}[!ht]
\caption{Summary of the $F$-test results.}
\smallskip
\begin{center}
{\small
\begin{tabular}{lcl}
\tableline
\noalign{\smallskip}
Dataset & Bandpass & Result \\         
\noalign{\smallskip}
\tableline
\noalign{\smallskip}
OGLE         & $B$ & Non-linear \\
OGLE         & $V$ & Non-linear \\
MACHO        & $V$ & Non-linear \\
MACHO        & $R$ & Non-linear \\
OGLE         & $I$ & Non-linear \\
OGLE + 2MASS & $J$ & Non-linear \\
MACHO + 2MASS& $J$ & Non-linear \\
OGLE + 2MASS & $H$ & Non-linear \\
MACHO + 2MASS& $H$ & Non-linear \\
OGLE + 2MASS & $K$ & (Marginally) Linear \\
MACHO + 2MASS& $K$ & (Marginally) Linear \\
OGLE + IRAS  &3.6$\mu$m& Linear \\
OGLE + IRAS  &4.5$\mu$m& Linear \\
OGLE + IRAS  &5.8$\mu$m& Linear \\
\noalign{\smallskip}
\tableline
\end{tabular}
}
\end{center}
\end{table}

In Table 1, we summarize the $F$-test results \citep[taken from][]{kan04,kan06,nge05,nge08a,nge08b} in different bandpasses from various datasets. The fact that the non-linear results were obtained from two different datasets, namely the OGLE (Optical Gravitational Lensing Experiment) and the MACHO (MAssive Compact Halo Objects) data, reinforce the existence of non-linear LMC P-L relations. Simple black-body arguments with Cepheid-like temperatures can be used to explain the results given in Table 1 \citep[see][for more details]{nge06b}. This suggests that Cepheid temperature (or equivalently the color) plays an important role in the observed non-linear P-L relation. In addition to the P-L relation, the period-color (P-C) relation and the instability strip were also found to be non-linear in the optical for the LMC Cepheids \citep{kan04,san04,kan06}.

\section{The HIF - Photosphere Interaction}

The hydrogen ionization front (HIF) is a region of rapid temperature change (together with a sharp rise of the opacity) near the surface of Cepheids, where hydrogen is partially ionized with a characteristic temperature. Due to its radial pulsation, the HIF will move in-and-out in mass distribution for a given Cepheid, and both of the HIF and photosphere are not co-moving. Therefore, at certain phases of the pulsation, the photosphere of the Cepheid (defined as a layer with optical depth of $2/3$) can be located at the base of HIF. The ``opacity wall'' from the HIF prevents the photosphere moving in further into the mass distribution. 

Depending on the period, metallicity and the density, the photosphere can interact with HIF when it is located at the base of HIF, while the density is reasonably low. In this situation, the temperature of the photosphere is the characteristic temperature of the HIF which ionizes hydrogen. A consequence of this HIF-photosphere interaction is that the P-C relation will be flatter (i.e., more independent of the pulsation period) for phases at which this interaction occurs at low densities. The flat P-C relation has been observed for long period Galactic and LMC Cepheids at the maximum light \citep{sim93,kan04,kan04a,kan06} and for RR Lyrae at the minimum light \citep{stu66,kan95,kan96,kan05}. Since the P-L and P-C relations are 2-dimensional projection of the P-L-C relation, and the P-L relations normally stated in the literature (such as those tested in Table 1) are {\it mean} light relations (i.e., average over the pulsation phases), the ``abnormality'' of the P-C relation at certain phases could lead to the observed non-linear mean light P-L relation. 

To test the effect of HIF - photosphere interaction on the P-C relation, we constructed sequences of stellar pulsation models appropriate for the Galactic \citep{kan04a} and the LMC \citep{kan06} Cepheids. From the models, we identified the locations of the photosphere and HIF in the mass distribution, $q=\log[1-M(r)/M]$, where $M(r)$ is the mass within radius $r$ and $M$ is the total mass of a given Cepheid model. A qualitative way to indicate the HIF-photosphere interaction is by plotting the ``distances'' between HIF and photosphere, in terms of $q$, against the pulsational periods as shown in Figure 15 of \citet{kan06}. From that Figure, the HIF is located near or at the base of the HIF at the maximum light for both of the Galactic and LMC models, which lead to the observed flat P-C relation at maximum light. However at minimum light, photospheres from the Galactic models are clearly far away from the HIF in mass distribution, and hence no interaction occur with the HIF. The LMC models, on the other hand, show that the photosphere is disengaged from the HIF at minimum light for models with period greater than 10 days. In contrast, the short period LMC models suggest the HIF - photosphere interaction occurs at most phases. This different behavior of the long and short period LMC models may be responsible for the observed non-linear LMC P-L relation at mean light. 

\section{Conclusion}

Using the $F$-test, LMC Cepheid P-L relations are found to be non-linear in the optical and near infrared $JH$ bands, and become linear around $K$ band and longer wavelength, as summarized in Table 1. These results, from a simple black-body argument, suggest temperature plays an important role in explaining the non-linear P-L relation. One mechanism that can affect the photospheric temperature at certain phases of the pulsation is the HIF - photosphere interaction. This interaction can produce a flatter P-C relation at maximum light as observed in the Galactic and LMC Cepheids. Our Cepheid models also imply that a distinct behavior between the Galactic and LMC models at various phases can lead to the observed non-linear LMC P-L relation at mean light. 

The discovery of the non-linearity in the LMC P-L relation has challenged the long standing assumption that the Cepheid P-L relation is linear. The entire Cepheid based distance scale is based on this assumption and the situation needs to be addressed. A more accurate ($<5\%$) measurement of $H_0$ than exists at present is crucial because it will help to break the degeneracy between ${\Omega}_M$ and $H_0$ in CMB measurements \citep[e.g.,][]{teg04,hu05,spe07}, required in the era of precision cosmology. The optimal way to do this is to calibrate the Cepheid distance scale to an accuracy of a few percent. \citet{nge06c} show that the error in estimating $H_0$ arising from using a linear P-L relation when the true relation is intrinsically non-linear is about 2\%. As other sources of larger systematic error in $H_0$ are addressed (e.g. the absolute calibration or the metallicity dependence) and a number of recent attempts at reducing zero point errors on the Cepheid distance scale \citep[e.g., see][]{mac06}, our lack of knowledge about the slope of the P-L relation will become increasingly important. Equally as important, a proper understanding of the non-linear P-L relation is vital for stellar evolution/pulsation studies of Cepheids, to develop a theory explaining the physics behind these observations. One such theory is the proposed HIF - photosphere interaction as described in previous section and in \citet{kan06}.

\acknowledgements 
C. N. acknowledges the support of the American Astronomical Society and the National Science Foundation in the form of an International Travel Grant, which enabled him to attend this conference.

\end{document}